\documentclass[12pt,a4paper]{article}%
\usepackage{amsmath,amssymb,amsfonts}
\usepackage{graphicx,epsfig}
\usepackage{hyperref}
\usepackage{color}
\usepackage{amsmath}
\usepackage{amsfonts}
\usepackage{amssymb}
\usepackage{graphicx}
\usepackage{slashed}%
\numberwithin{equation}{section}
\newcommand{\be}{\begin{equation}}
\newcommand{\ee}{\end{equation}}
\newcommand{\bea}{\begin{eqnarray}}
\newcommand{\eea}{\end{eqnarray}}
\newcommand{\ba}{\begin{array}}
\newcommand{\ea}{\end{array}}

\newcommand{\cO}{{\cal O}}

\definecolor{red}{cmyk}{0,1,1,0.4}
\definecolor{blu}{cmyk}{1,1,0,0.3}
\definecolor{green}{cmyk}{0.92,0,0.59,0.25}
\begin{document}
\begin{titlepage}
\begin{flushright}
\end{flushright}
\vskip 1.0cm
\begin{center}
{\Large \bf Heavy Vectors in Higgs-less models} \vskip 1.0cm
{\large Riccardo Barbieri$^a$,\ Gino Isidori$^{a,b}$,\ Vyacheslav S.~Rychkov$^a$,\ Enrico Trincherini$^a$} \\[1cm]
{\it $^a$ Scuola Normale Superiore and INFN, Piazza dei Cavalieri 7, 56126 Pisa, Italy} \\[5mm]
{\it $^b$ INFN, Laboratori Nazionali di Frascati, Via E. Fermi 40
I-00044 Frascati, Italy}\\[5mm]
\vskip 1.0cm \abstract{One or more heavy spin-1 fields may  {\it
replace} the Higgs boson in keeping perturbative unitarity up to a
few TeV. By means of two prototype  chiral models for the heavy
spin-1 bosons, a {\it composite} model or a {\it gauge} model, we
discuss if and how the sole exchange of the same fields can also
account for the ElectroWeak Precision Tests. While this proves
impossible in the gauge model, the composite model hints to a
positive solution, which we exploit to constrain the
phenomenological properties of the heavy vectors. }
\end{center}
\end{titlepage}

\section{Introduction and statement of the problem}

\label{sect:intro}

The case for the existence of a (relatively light) Higgs boson is strong. In
the Standard Model (SM) all its couplings are determined in terms of a single
parameter, its mass $m_{h}$. By taking $m_{h}$ around 100~GeV, all the
ElectroWeak Precision Tests (EWPT) are successfully accounted for. With the
inclusion of the Higgs boson the electroweak interactions are consistently
described up to energies which are, in principle, indefinitely high. Yet the
strong sensitivity of $m_{h}$ to short distance scales makes it hard to
understand its relative lightness and motivates the search for alternative
roads. Most of the times one considers an extended Higgs sector, in one
direction or another. Relatively less frequently, and generally less
successfully, one also tries to do without it at all.

In this last case one or more spin-1 bosons often \textit{replace} the Higgs
boson in trying to keep perturbative unitarity up to a few TeV, one order of
magnitude above the Fermi scale, $v=(\sqrt{2}G_{F})^{-1/2}\simeq246$~GeV, so
that a meaningful comparison with current experiments can be made. While this
is possible, the role played by the Higgs boson of the SM in the EWPT proves
generally harder to substitute, at least without adding extra ingredients that
have little to do with the spin-1 bosons themselves. This is a serious
drawback. Not only the competition with the Higgs boson gets weakened but
also, unlike in the Higgs boson case, one cannot use the EWPT to determine the
very properties of the heavy vector(s) which are phenomenologically crucial.
This is the point that we examine in this work by means of two prototype
chiral models for the heavy spin-1 bosons: a \textit{composite} model and a
\textit{gauge} model. As we are going to see, while the gauge model confirms
the difficulty, the composite model hints to a possible positive solution,
which we exploit to constrain the phenomenological properties of the heavy
vectors. The corresponding effective Lagrangians are meant to be valid up a
cutoff $\Lambda\approx4\pi v\approx3$~TeV.

In connection with the EWPT, a much discussed problem is the effect on the
$S$-parameter of the tree level exchange of the heavy spin-1 fields. We shall
assume that $S$ at tree level is positive and moderately small,
$0<S^{\text{tree}}\lesssim0.2$, as compatible with both models in a suitable
range of their parameters. At least equally important, however, is the
correlation of $S$ with $T$, which, although vanishing at tree level in
presence of a suitable custodial symmetry, has a one loop \textit{infrared}
effect
\begin{equation}
\left.  \hat{T}\right\vert _{\text{IR}}=-\frac{3\alpha}{8\pi c_{W}^{2}}%
\ln\left(  \frac{\Lambda}{m_{W}}\right)  =-1.2\cdot10^{-3}\ln\left(
\frac{\Lambda}{m_{W}}\right)  , \label{Tinfrared}%
\end{equation}
which in the SM is cut off by the Higgs boson exchange. By itself, with
$\Lambda\approx3$~TeV, it would be largely incompatible with current data, no
matter what happens to $S$.
On the other hand the
exchange of the heavy vectors in the loop are bound to give a contribution
which may mitigate this effect, as they have couplings to pions delaying the
loss of unitarity in $WW$-scattering. If one wants to defend the calculability
of the EWPT, the full one loop for $T$ must therefore be included. As we will
see, the compensating effect of the heavy vectors is more subtle than the one
of the Higgs boson; in particular it does not reduce to replacing the
logarithmic cutoff $\Lambda$ in (\ref{Tinfrared}) with the heavy vector mass.

The common starting point of all Higgs-less models in the literature is the
$SU(2)_{L}\times SU(2)_{R}$ \textit{chiral} symmetry spontaneously broken down
to $SU(2)_{L+R}$. A minimal point of view consists therefore in taking the
heavy spin-1 field(s) as (triplet) non-linear representation(s) of
$SU(2)_{L}\times SU(2)_{R}/SU(2)_{L+R}$. The $W$ and $B$ bosons
\textit{weakly} gauge the usual $SU(2)_{L}\times U(1)_{Y}$ subgroup, with $Y$
extended to include $B-L$ when acting on fermions, which only couple to $W$
and $B$. While not crucial to our conclusions, for simplicity we shall assume
parity conservation as the usual gauge couplings, $g$ and $g^{\prime}$, are
switched off. We shall therefore speak of vector or axial spin-1 fields. This
essentially defines the \textit{composite} model.

Although with some constraints, as we are going to see, a structure of this
type may emerge in particular from a gauge theory based on $G=SU(2)_{L}\times
SU(2)_{R}\times SU(2)^{N}$ broken to the diagonal subgroup
$H=SU(2)_{L+R+\ldots}$ by a generic non-linear $\sigma$-model of the form
\begin{equation}
\mathcal{L}_{\chi}=\sum_{I,J}v_{IJ}^{2}\langle D_{\mu}\Sigma_{IJ}D^{\mu}%
\Sigma_{IJ}^{\dagger}\rangle~,\qquad\Sigma_{IJ}\rightarrow g_{I}\Sigma
_{IJ}g_{J}^{\dagger}~, \label{eq:one}%
\end{equation}
where $g_{I,J}$ are elements of the various $SU(2)$ and $D_{\mu}$ are
covariant derivatives of $G$. This is the \textit{gauge} model that we shall
consider. It includes as special cases or approximates via deconstruction many
of the models in the
literature~\cite{Casalbuoni:1985kq,Csaki:2003dt,Foadi:2003xa,Georgi:2004iy}.

The paper is organized as follows. In Sect. 2 we give the Lagrangian of the
composite model. In Sect. 3 we recall the unitarity constraints. In Sect. 4 we
calculate the contributions to $T$ of the heavy vector exchanges in the
composite model. In Sect. 5 we do the analogous calculation in the gauge model
and discuss the relation between the two models. In Sect. 6 we summarize the
properties of the heavy spin-1 field(s) in the composite model as they emerge
from the unitarity and the EWPT constraints. The corresponding phenomenology
is briefly described in Sect. 7. Conclusions are drawn in Sect. 8. In the
Appendices we discuss the sum rules following from the assumption of spin-1
meson dominance, and review the phenomenological parameters of low-lying
spin-1 resonances in QCD.

\section{The Lagrangian of the \textit{composite} model}

\label{sect:compo}

The building block is the usual lowest-order chiral Lagrangian for the
Goldstone fields
\begin{equation}
\mathcal{L}_{\chi}^{(2)}(U)=\frac{v^{2}}{4}\langle D_{\mu}U(D^{\mu}%
U)^{\dagger}\rangle~, \label{eq:L2}%
\end{equation}
where
\begin{align}
&  U=e^{i2\hat{\pi}/v},\qquad\hat{\pi}=T^{a}\pi^{a}=\frac{1}{\sqrt{2}}\left[
\begin{array}
[c]{cc}%
\frac{\pi^{0}}{\sqrt{2}} & \pi^{+}\nonumber\\
\pi^{-} & -\frac{\pi^{0}}{\sqrt{2}}%
\end{array}
\right]  ~,\qquad T^{a}=\frac{1}{2}\sigma^{a},\nonumber\\
&  D_{\mu}U=\partial_{\mu}U-i\hat{B}_{\mu}U+iU\hat{W}_{\mu}~,\qquad\hat
{W}_{\mu}=gT^{a}W_{\mu}^{a}~,\qquad\hat{B}_{\mu}=g^{\prime}T^{3}B_{\mu}~,
\label{def}%
\end{align}
and $\langle\rangle$ denotes the trace of a $2\times2$ matrix. The invariant
kinetic and mass terms for the standard fermions are left
understood. Under $SU(2)_{L}\times SU(2)_{R}$,
\begin{equation}
U\rightarrow g_{R}Ug_{L}^{\dagger}~.\nonumber
\end{equation}

Starting from $O(p^{4})$, there are many possible terms which can be added to
(\ref{eq:L2}). It is well known \cite{Ecker:1989yg} that in QCD the
coefficients of these terms can be near-saturated by the tree-level exchanges
of the low-lying spin-1 resonances (see App.~\ref{sect:QCD}). Motivated by
this fact, we will assume that this \textit{spin-1 meson dominance} is
approximately true in our composite model as well. As we will see, the
resulting framework is quite predictive.

Following Ref.~\cite{Ecker:1989yg}, we describe the heavy spin-1 states by
means of antisymmetric tensors. This formalism is particularly convenient
since it avoids any mixing of the spin-1 fields with (the derivatives of) the
Goldstone fields.
Up to field redefinitions, addition of local terms, and appropriate matching
conditions for the coupling constants, the results are equivalent to those
obtained with the more familiar formalism of vector fields~\cite{Ecker:1989yg}.

We consider at most two sets of vector states, $A^{\mu\nu}$ and $V^{\mu\nu}$,
with opposite parity, both transforming in the adjoint representation of
$SU(2)_{L+R}$
\begin{equation}
R^{\mu\nu}\rightarrow hR^{\mu\nu}h^{\dagger}~,\qquad R^{\mu\nu}=A^{\mu\nu
},\ V^{\mu\nu}~. \label{eq:Rtr}%
\end{equation}
To describe their transformation properties under the full $SU(2)_{L}\times
SU(2)_{R}$, we introduce the \textit{little} matrix $u$ \cite{coleman} via%
\begin{equation}
U=u^{2}~.\nonumber
\end{equation}
This matrix parametrizes the $SU(2)_{L}\times SU(2)_{R}/SU(2)_{L+R}$ coset and
transforms as%
\begin{equation}
u\rightarrow g_{R}uh^{\dagger}=hug_{L}^{\dagger}~,\nonumber
\end{equation}
where $h=h(u,g_{L},g_{R})$ is uniquely determined by this equation. The
general transformation of $R^{\mu\nu}$ is then given by the same Eq.
(\ref{eq:Rtr}) with $h$ so defined. For $g_{L}=g_{R}$ we have $h=g_{L}=g_{R}$
independent of $u$, and we recover the linear $SU(2)_{L+R}$ transformation
\cite{coleman}.

The kinetic Lagrangian for heavy spin-1 fields has the form:
\begin{equation}
\mathcal{L}_{\mathrm{kin}}(R^{\mu\nu})=-\frac{1}{2}\langle\nabla_{\mu}%
R^{\mu\nu}\nabla^{\sigma}R_{\sigma\nu}\rangle+\frac{1}{4}M_{R}^{2}\langle
R^{\mu\nu}R_{\mu\nu}\rangle~,
\end{equation}
where the covariant derivative
\begin{equation}
\nabla_{\mu}R=\partial_{\mu}R+[\Gamma_{\mu},R],\qquad\Gamma_{\mu}=\frac{1}%
{2}\left[  u^{\dagger}(\partial_{\mu}-i\hat{B}_{\mu})u+u(\partial_{\mu}%
-i\hat{W}_{\mu})u^{\dagger}\right]  ,\quad\Gamma_{\mu}^{\dagger}=-\Gamma_{\mu
},
\end{equation}
ensures that $\nabla_{\mu}R$ transforms as $R$ under the global $SU(2)_{L}%
\times SU(2)_{R}$ and under the SM gauge group.

The most general $SU(2)_{L}\times SU(2)_{R}$ invariant Lagrangian
at $\mathcal{O}(p^{2})$ describing the coupling of these heavy fields to
Goldstone bosons and SM gauge fields, invariant under parity, is
\begin{align}
&  \mathcal{L}_{V}(R,u)=\mathcal{L}_{\mathrm{kin}}(A^{\mu\nu})+\mathcal{L}%
_{\mathrm{kin}}(V^{\mu\nu})+\frac{i}{2\sqrt{2}}G_{V}\langle V^{\mu\nu}[u_{\mu
},u_{\nu}]\rangle\nonumber\\
&  \qquad+\frac{1}{2\sqrt{2}}F_{V}\langle V^{\mu\nu}(u\hat{W}^{\mu\nu
}u^{\dagger}+u^{\dagger}\hat{B}^{\mu\nu}u)\rangle+\frac{1}{2\sqrt{2}}%
F_{A}\langle A^{\mu\nu}(u\hat{W}^{\mu\nu}u^{\dagger}-u^{\dagger}\hat{B}%
^{\mu\nu}u)\rangle~, \label{eq:LV}%
\end{align}
where
\begin{equation}
u_{\mu}=iu^{\dagger}D_{\mu}Uu^{\dagger}=u_{\mu}^{\dagger},~\qquad u_{\mu
}\rightarrow hu_{\mu}h^{\dagger}~.
\end{equation}
The phenomenological parameters $G_{V},F_{V,A}$ have dimension of mass
and, by naive dimensional analysis, we expect them to be of $\cO(v)$.

The Lagrangian (\ref{eq:LV}) does not include any trilinear coupling between
the Goldstone fields and a pair of heavy spin-1 fields. We shall have to come
back to this point.

\section{Perturbative unitarity}

\label{sect:uni}

Up to terms of order $m_{W}/\sqrt{s}$, the amplitudes for longitudinal
gauge-boson scattering are identical to those of the corresponding Goldstone
bosons. In the limit of unbroken $SU(2)_{L+R}$ symmetry we can decompose the
generic $W_{L}^{a}W_{L}^{b}\rightarrow W_{L}^{c}W_{L}^{d}$ amplitude
as~\cite{Bagger}
\begin{equation}
\mathcal{A}(W_{L}^{a}W_{L}^{b}\rightarrow W_{L}^{c}W_{L}^{d})=\mathcal{A}%
(s,t,u)\delta^{ab}\delta^{cd}+\mathcal{A}(t,s,u)\delta^{ac}\delta
^{bd}+\mathcal{A}(u,t,s)\delta^{ad}\delta^{bc}~.
\end{equation}
The fixed isospin amplitudes are~\cite{Bagger}
\begin{align*}
T(0)  &  =3A(s,t,u)+A(t,s,u)+A(u,t,s)\\
T(1)  &  =A(t,s,u)-A(u,t,s)\\
T(2)  &  =A(t,s,u)+A(u,t,s)
\end{align*}
and the partial wave coefficients:
\begin{align*}
a_{l}^{I}(s)  &  =\frac{1}{64\pi}\int_{-1}^{1}d(\cos\theta)P_{l}(\cos
\theta)T(I)\,,\\
t  &  =-\frac{s}{2}(1-\cos\theta),\quad u=-\frac{s}{2}(1+\cos\theta)\,.
\end{align*}

Evaluating this process at the tree level leads to
\begin{equation}
\mathcal{A}(s,t,u) = i\mathcal{A}(\pi^{+}\pi^{-} \to\pi^{0}\pi^{0}) = \frac
{s}{v^{2}} - \frac{G^{2}_{V}}{v^{4}} \left[  3 s + M_{V}^{2} \left(  \frac{s-u
}{t-M_{V}^{2}} + \frac{s-t}{u-M_{V}^{2}}\right)  \right]  ~, \label{eq:App}%
\end{equation}
where the first term is the contribution of $\mathcal{L}_{\chi}^{(2)}$.
Exchanges of more than one vector spin-1 field are easily included. Axial
spin-1 fields do not contribute to $\mathcal{A}(s,t,u)$.

The cancellation in (\ref{eq:App}) of the linear growth with $s$ occurs for
\begin{equation}
G_{V}=v/\sqrt{3}~.
\end{equation}
This result is equivalent to the one obtained by Bagger \textit{et
al.}~\cite{Bagger} ($a=4G_{V}^{2}/v^{2}$ in their notation). The strongest
unitarity constraint is obtained by requiring $|a_{0}^{0}|<1$ for any energy
up to $\sqrt{s}=\Lambda$, where
\begin{equation}
a_{0}^{0}=\frac{M_{V}^{2}}{16\pi v^{2}}\left\{  x\left(  1-\frac{3G_{V}^{2}%
}{v^{2}}\right)  +\frac{2G_{V}^{2}}{v^{2}}\left[  (2+x^{-1})\log
(x+1)-1\right]  \right\}  ~,\qquad x=\frac{s}{M_{V}^{2}}~.
\end{equation}

\begin{figure}[t]
\begin{center}
\includegraphics[width=12cm]{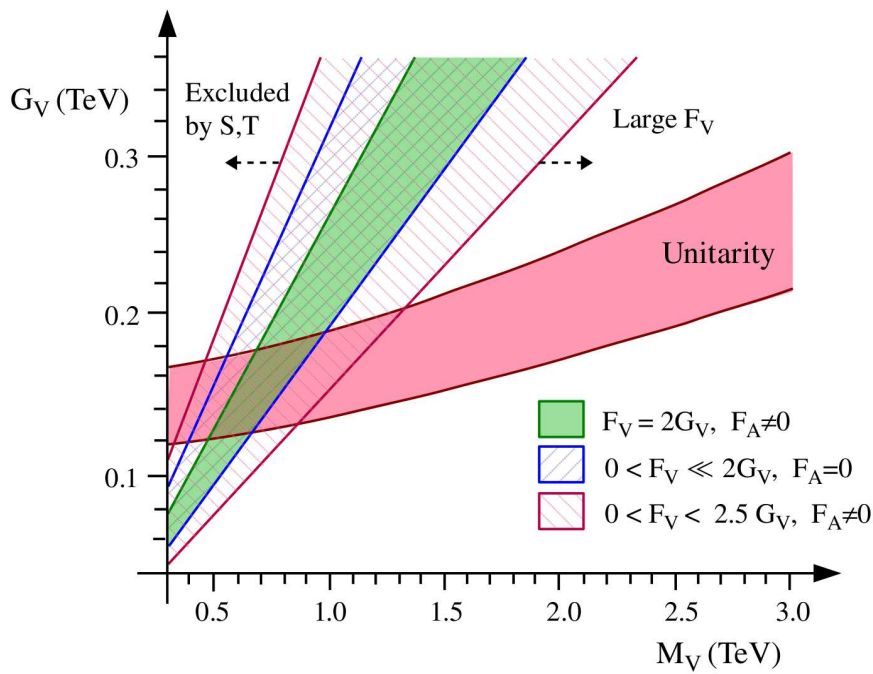}
\end{center}
\caption{Summary of unitarity and EWPT constraints (at 95\% C.L.) in the
$(M_{V},G_{V})$ plane (see Sect.~\ref{sect:constr}). }%
\label{fig:GVMV}%
\end{figure}

Imposing the constraint $|a_{0}^{0}|<1$ up to $\Lambda\simeq3$ TeV, we get an
allowed region in the $(M_{V},G_{V})$ plane extending all the way up in
$M_{V}$, see Fig.~\ref{fig:GVMV}. Notice that the partial wave grows with
energy even for $G_{V}=v/\sqrt{3}$, albeit only logarithmically. As a result
the unitarity is restored more efficiently for $G_{V}$ moderately above
$v/\sqrt{3}$, so that the linear term enters with a small negative coefficient
and compensates for the logarithmic growth. This is the situation realized for
the $\rho$ meson in QCD (see App.~\ref{sect:QCD}).\ Of course this mechanism
cannot work for arbitrarily high energies, also because at higher energies we
have to take inelastic channels into account~\cite{Papucci}.

It is interesting to note that models of electroweak symmetry breaking in 5D
and their $N$-site deconstructions, to be discussed in Sect. \ref{sect:gauge},
apparently cannot access the region $G_{V}\geq v/\sqrt{3}$. For instance in
the 3-site model~\cite{Matsuzaki:2006wn} one has $G_{V}=v/2$, so that the
vector exchange cancels only $3/4$ of the linear $s$ growth of the amplitude.
In those models the linear growth can be canceled only by the exchange of the
\textit{full} tower of resonances, i.e.~in 5D or in the limit of infinitely
many sites, and even then the amplitude continues to grow logarithmically.

The analogue of the amplitude (\ref{eq:App}) in the SM with
the Higgs boson exchange is 
\begin{align*}
&  A(s,t,u)=-\frac{M_{H}^{2}}{v^{2}}\frac{s}{s-M_{H}^{2}}\,,\\
&  a_{0}^{0}=\frac{1}{16\pi}\frac{M_{H}^{2}}{v^{2}}\left(  \frac{\log(x+1)}%
{x}-\frac{3/2}{x-1}-5/2\right)  ~,\qquad x=\frac{s}{M_{H}^{2}}~.
\end{align*}
This partial wave has a fixed limit at large $s$, which however grows with
$M_{H}$. Because of this, unlike in the vector-boson case, $M_{H}>1.2$ TeV is
not compatible with the unitarity bound~\cite{Lee}.

Finally, we remind that the loss of unitarity associated with the chiral
Lagrangian description of fermion masses intervenes at energies 
well above 3 TeV.

\section{$T$ at one loop in the composite model}

\label{sect:T}

\begin{figure}[t]
\begin{center}
\includegraphics[width=15cm]{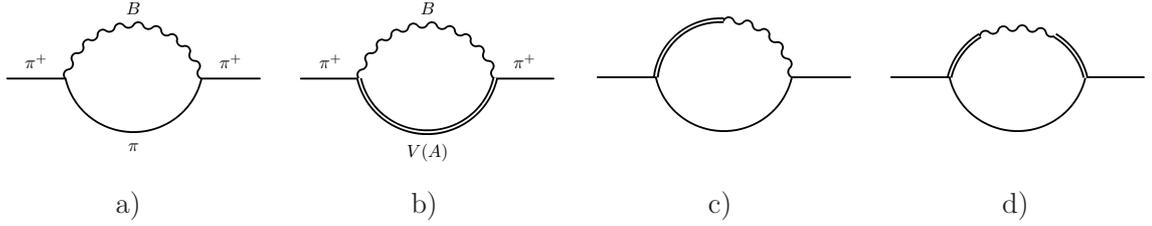}
\end{center}
\caption{One-loop contributions to $\hat{T}$ generated by $\mathcal{L}^{(2)}$
and by couplings of the heavy vectors at most linear in the heavy fields. }%
\label{fig:T1loop}%
\end{figure}

The tree level exchange of the heavy vectors leads to well known contributions
to the parameters $\hat{S}, W, Y$ of Ref.~\cite{Barbieri:2004qk}:%

\begin{equation}
\Delta\hat{S}=g^{2}\left(  \frac{F_{V}^{2}}{4M_{V}^{2}}-\frac{F_{A}^{2}%
}{4M_{A}^{2}}\right)  ~,\qquad\Delta W=\left(  \frac{g}{g^{\prime}}\right)
^{2}\Delta Y=g^{2}m_{W}^{2}\left(  \frac{F_{V}^{2}}{4M_{V}^{4}}+\frac
{F_{A}^{2}}{4M_{A}^{4}}\right)  ~,
\end{equation}
whereas it leaves $T$ untouched because of the protecting $SU(2)_{L+R}$
symmetry. As already pointed out, however, it is crucial to include the loop
effects in T, which arise from the diagrams of Fig.~\ref{fig:T1loop}. The
corresponding calculation is in fact significantly simplified by noticing
that, up to corrections of relative order $m_{W}^{2}/M_{R}^{2}$,
\begin{equation}
\hat{T}=\frac{Z^{(+)}}{Z^{(0)}}-1 \label{eq:TZ}%
\end{equation}
where $Z^{(+)},Z^{(0)}$ are the wave-function renormalization constants of the
charged and neutral Goldstone bosons computed \textit{in the Landau gauge} for
the light vectors. This follows from the Ward identities of the global
$SU(2)_{L}$ symmetry~\cite{Barbieri:1992dq}
and from the fact that it is only in the Landau gauge that the global
$SU(2)_{L}\times U(1)_{Y}$ is preserved and the Goldstone bosons are kept
massless. A further simplification occurs by setting to zero the $SU(2)_{L}$
gauge coupling, which does not break the custodial symmetry, so that only the
B-boson exchange produces an effect in $T$.

From the diagram of Fig.~\ref{fig:T1loop}a, which is there in the SM, one
obtains the infrared effect in (\ref{Tinfrared}). The contribution from the
remaining diagrams is also readily obtained by means of the propagator for the
spin-1 fields of mass $M$, vector or axial, in the antisymmetric tensor
paramatrization that we are using \cite{Ecker:1989yg}
\begin{equation}
D_{\mu\nu,\rho\sigma}(k)=i\frac{g_{\mu\rho}g_{\nu\sigma}-g_{\nu\rho}%
g_{\mu\sigma}}{k^{2}-M^{2}}-i\frac{P_{\mu\nu,\rho\sigma}(k)}{k^{2}-M^{2}},
\label{eq:prop}%
\end{equation}%
\begin{equation}
P_{\mu\nu,\rho\sigma}(k)=-g_{\mu\rho}g_{\nu\sigma}k^{2}-k_{\mu}k_{\sigma
}g_{\nu\rho}+k_{\mu}k_{\rho}g_{\nu\rho}-(\mu\leftrightarrow\nu).
\end{equation}
Except for special values of the couplings, the dominant contribution to $T$
from the heavy spin-1 exchanges comes from the diagram of
Fig.~\ref{fig:T1loop}b, which is the only one to diverge
quadratically\footnote{The remaining two diagrams do not contain a quadratic
divergence because the second term in (\ref{eq:prop}), which dominates at high
momenta, vanishes when contracted with $k_{\mu}$ in any of the four indices.}.
Adding this contribution to the infrared term we have
\begin{equation}
\left.  \Delta\hat{T}\right\vert _{\mathrm{generic}}=-\frac{3\alpha}{8\pi
c_{W}^{2}}\ln\left(  \frac{\Lambda}{m_{W}}\right)  +\frac{3\pi\alpha}%
{c_{W}^{2}}\left[  \frac{F_{A}^{2}}{4M_{A}^{2}}+\left(  \frac{F_{V}-2G_{V}%
}{2M_{V}}\right)  ^{2}\right]  \frac{\Lambda^{2}}{16\pi^{2}v^{2}}%
+\ldots\label{Tlambda2}%
\end{equation}

To make contact with the gauge model as defined in the Introduction and with
the existing literature on the subject, it is useful to consider the same
calculation of $T$ for the special case $F_{A}=0,F_{V}=2G_{V}$, which makes
the $\Lambda^{2}$-term in (\ref{Tlambda2}) to vanish. In this case all the
four diagrams in Fig.~\ref{fig:T1loop} contribute with a logarithmic term or,
explicitly,
\begin{equation}
\left.  \Delta\hat{T}\right\vert _{F_{V}=2G_{V},F_{A}=0}=-\frac{3\alpha}{8\pi
c_{W}^{2}}\left\{  \ln\left(  \frac{M_{V}}{m_{W}}\right)  +\left[  \left(
1-2\frac{G_{V}^{2}}{v^{2}}\right)  ^{2}+\frac{G_{V}^{2}}{v^{2}}\right]
\ln\left(  \frac{\Lambda}{M_{V}}\right)  +\mathcal{O}(1)\right\}
\label{eq:Tlogs}%
\end{equation}
The result of Ref.~\cite{Matsuzaki:2006wn} is recovered from (\ref{eq:Tlogs})
for $G_{V}=v/2$. This is a particular 3-site gauge model with two links
connecting $SU(2)_{L}$ and $SU(2)_{R}$ to one extra $SU(2)$ respectively. One
can obtain $G_{V}<v/2$ in this model, still preserving parity, by having one
extra link connecting $SU(2)_{L}$ and $SU(2)_{R}$ with each other. However,
for any choice of $G_{V}/v$ the vector contribution does not compensate the
dominant infrared term.

Two interesting questions arise at this point. In the QCD case, where the
Lagrangian (\ref{eq:LV}) is known to describe well the properties of the
spin-1 resonances and, especially, their contribution to the low energy
pion-interactions, is there any evidence for $F_{A} \ne0$ and/or $F_{V}\ne2
G_{V}$? Furthermore, if indeed $F_{A} \ne0$ and/or $F_{V}\ne2 G_{V}$, what
cuts off the quadratic divergence in (\ref{Tlambda2})?

To the first question there is a neat answer. By integrating out the heavy
spin-1 fields at tree level, one shows (See Sect. \ref{sect:gauge}) that a
trilinear coupling $\pi-v_{\mu}(q)-a_{\nu}(k)$ arises between the pion field
and the external vector and axial currents carrying respectively momentum $q$
and $k$. Furthermore this coupling at $q^{2}=k^{2}=0$ is non vanishing if and
only if $F_{A}\neq0$ and/or $F_{V}\neq2G_{V}$~(see App.~\ref{sect:QCD}).
Precise measurements of the radiative pion decay, $\pi^{+}\rightarrow e^{+}%
\nu\gamma$, show that this same coupling is phenomenologically
required~\cite{PIBETA}. A non-vanishing $F_{A}$ is also required 
by the significant partial width $\Gamma(a_{1}\rightarrow\pi\gamma$)
(see App.~\ref{sect:QCD}).

\begin{figure}[t]
\begin{center}
\includegraphics[width=8.5cm]{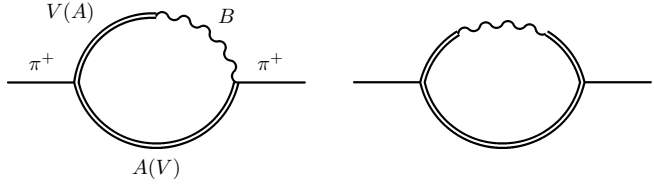} \vskip  0.5 cm
\end{center}
\caption{One-loop contributions to $\hat T$ generated by the $A V \pi$
trilinear couplings.}%
\label{fig:Tcutoff}%
\end{figure}

We will therefore assume that, in analogy to QCD, $F_{A}\neq0$ and/or
$F_{V}\neq2G_{V}$, and try to take advantage of the positive second term in
(\ref{Tlambda2}). In order to convince ourselves that this quadratically
divergent term corresponds to a true physical effect, it is important to
understand a mechanism which cuts off the quadratic divergence. This would
simultaneously provide us with an idea for the appropriate $\Lambda$. The
mechanism that we propose involves the trilinear interactions between the
Goldstone pions and a pair of heavy spin-1 fields
\begin{equation}
\mathcal{L}_{2V}(R,u)=ig_{j}^{A}\langle A^{\mu\nu}[\nabla_{\rho}V_{j}^{\rho
\nu},u_{\mu}]\rangle+ig_{j}^{V}\langle V^{\mu\nu}[\nabla_{\rho}A_{j}^{\rho\nu
},u_{\mu}]\rangle~.\label{eq:L2V}%
\end{equation}
These couplings are required if one imposes the vanishing at high momentum of
the form factors
\begin{equation}
\langle A|v_{\mu}(q)|\pi\rangle~,\qquad\qquad\langle V|a_{\mu}(q)|\pi
\rangle~,\label{eq:formfactors}%
\end{equation}
for each vector or axial field. As explicitly indicated with the index $j$,
this may involve, in general, more than one heavy state. In App.~A we discuss
explicitly the constraints of this type, as well as those arising from the
high momentum behaviour of the pion form factor and of the left-right
two-point function $(\langle v_{\mu}v_{\nu}\rangle-\langle a_{\mu}a_{\nu
}\rangle)$ (the analogue of the Weinberg sum rules in QCD). From the couplings
in (\ref{eq:formfactors}) more diagrams contribute to $T$ with two heavy
spin-1 particles in the intermediate states (see Fig.~\ref{fig:Tcutoff}). With
the form factors in (\ref{eq:formfactors}) softened by the couplings
(\ref{eq:L2V}), the quadratic divergence in (\ref{Tlambda2}) disappears and
$\Lambda$ gets replaced by a heavy particle mass. We will assume that this
cancellation involves heavy states of mass close to the cutoff, so that
(\ref{Tlambda2}) continues to be a good estimate of $T$ with $\Lambda\approx3$~TeV.

\section{The $SU(2)^{N}$ gauge model}

\label{sect:gauge}

As anticipated, we consider a generic model based on $G=SU(2)_{L}\times
SU(2)_{R}\times SU(2)^{N}$, with the $SU(2)^{N}$ subgroup fully gauged and
with $G$ spontaneously broken into the diagonal subgroup $H$. The Lagrangian
of the model is
\begin{equation}
\mathcal{L}=\mathcal{L}_{\mathrm{gauge}}+\mathcal{L}_{\chi},\quad
\mathcal{L}_{\mathrm{gauge}}=-\sum_{I}\frac{1}{4g_{I}^{2}}(\omega_{I}^{\mu\nu
})^{2},\quad\omega_{I}^{\mu\nu}=\partial^{\mu}\omega_{I}^{\nu}-\partial^{\nu
}\omega_{I}^{\mu}-i[\omega_{I}^{\mu},\omega_{I}^{\nu}]~, \label{eq:Lgauge}%
\end{equation}
with $\mathcal{L}_{\chi}$ given in (\ref{eq:one}).

This model has been analyzed in \cite{Georgi:2004iy} and argued to give
$\Delta S_{\text{tree}}\gtrsim N/6\pi$, in accordance with a previous result
in 5D \cite{Barbieri:2003pr}. Here we are mainly interested in the relation of
this model with the composite model of Section \ref{sect:compo}. To this end
we analyze its low energy limit after the tree level integration of the heavy
vectors and we compute $T$ at one loop. Unlike the case of the composite
model, we show that the gauge model does not have an on shell $\pi-v_{\mu
}-a_{\nu}$ coupling of $\mathcal{O}(p^{4})$ and that it cannot account for the
EWPT by heavy vector exchanges only. The reader not interested in the gauge
model or in its connection with the composite model can skip this Section.

The symmetry breaking term of the model can be put in the form%
\begin{equation}
\mathcal{L}_{\chi}=\sum_{I,J}v_{IJ}^{2}\langle(\Omega_{I}^{\mu}-\Omega
_{J}^{\mu})^{2}\rangle~, \label{eq:non-diag}%
\end{equation}
where%
\[
\Omega_{I}^{\mu}=\sigma_{I}^{\dagger}\omega_{I}^{\mu}\sigma_{I}+i\sigma
_{I}^{\dagger}\partial^{\mu}\sigma_{I}~
\]
are gauge transformations of the original fields, and $\sigma_{I}$ are the
elements of $SU(2)_{I}/H$ (the generalizations of \emph{little} $u$:
$\sigma_{I}\overset{G}{\longrightarrow}g_{I}\sigma_{I}h^{\dagger}$). The link
fields can be expressed in terms of $\sigma_{I}$ as:
\[
\Sigma_{AB}=\phi_{AB}\sigma_{A}\sigma_{B}^{\dagger},
\]
where $\phi_{AB}$ is singlet under $G$ describing extra Higgs-like degrees of
freedom. We will assume In what follows that these extra scalars are
decoupled, $\phi_{AB}\equiv1$. This can be achieved by adding to the
Lagrangian (\ref{eq:Lgauge}) placket-like mass terms $\left\langle \Sigma
_{Ai}\ldots\Sigma_{jB}\Sigma_{AB}^{\dagger}\right\rangle $ for each
non-sequential link $\Sigma_{AB}$ and sending these masses to infinity.

We look for a description in terms of the SM gauge fields and the
$SU(2)_{L}\times SU(2)_{R}$ \ Goldstone $u$ defined via
\[
\sigma_{R}=\sigma_{L}^{\dagger}=:u\,.
\]
We hide the Goldstone boson fields of the non-SM groups in the heavy
$\Omega_{I\not =L,R}^{\mu}$, which corresponds to a partial gauge-fixing.
These fields transform under $SU(2)_{L}\times SU(2)_{R}$ via%
\[
\Omega_{I}^{\mu}\rightarrow h\Omega_{I}^{\mu}h^{\dagger}+ih\partial^{\mu
}h^{\dagger}~.
\]

The $\Omega_{I}^{\mu}$ can be decomposed with respect to parity
as\footnote{Note that $A_{I}^{\mu}\equiv0$ if $I=P(I).$}
\begin{align}
\Omega_{I}^{\mu}  &  =V_{I}^{\mu}+A_{I}^{\mu}~,\quad\Omega_{P(I)}^{\mu}%
=V_{I}^{\mu}-A_{I}^{\mu}\\
V_{I}^{\mu}  &  \rightarrow hV_{I}^{\mu}h^{\dagger}+ih\partial^{\mu}%
h^{\dagger},\quad A_{I}^{\mu}\rightarrow hA_{I}^{\mu}h^{\dagger}~
\label{eq:VAtrans}%
\end{align}
Defining
\begin{equation}
V_{K}^{\mu\nu}=\partial^{\mu}V_{K}^{\nu}-\partial^{\nu}V_{K}^{\mu}%
-i[V_{K}^{\mu},V_{K}^{\nu}]~,\qquad D_{V}^{\mu}A_{K}^{\nu}=\partial^{\mu}%
A_{K}^{\nu}-i[V_{K}^{\mu},A_{K}^{\nu}]~,
\end{equation}
we can write the gauge Lagrangian in the form:
\begin{equation}
\mathcal{L}_{\mathrm{gauge}}=\mathcal{L}_{\text{gauge,SM}}-\sum_{I\neq
L,R}\frac{1}{4g_{I}^{2}}\left[  \langle\left(  V_{I}^{\mu\nu}-i[A_{I}^{\mu
},A_{I}^{\nu}]\right)  ^{2}\rangle+\langle\left(  D_{V}^{\mu}A_{I}^{\nu}%
-D_{V}^{\nu}A_{I}^{\mu}\right)  ^{2}\rangle\right]  ~. \label{eq:Lgaugef}%
\end{equation}

Coming back to the symmetry breaking term, the quadratic form
(\ref{eq:non-diag}) can be brought by a field rotation to a diagonal form in
terms of mass eigenstates of different parity\footnote{This transformation can
be taken orthogonal in the vector sector. In the axial sector the orthogonal
transformation has to be followed by a non-orthogonal linear transformation
needed to separate the pure $\langle(u^{\mu})^{2}\rangle$ term.}:
\begin{align}
\mathcal{L}_{\chi}  &  =\sum_{n=1}^{N_{V}}v_{Vn}^{2}\langle(\hat{V}_{n}^{\mu
})^{2}\rangle+\sum_{n=1}^{N_{A}}v_{An}^{2}\langle(\hat{A}_{n}^{\mu}%
)^{2}\rangle+\frac{v^{2}}{4}\langle(u^{\mu})^{2}\rangle\label{eq:sep}\\
\hat{V}_{n}^{\mu}  &  =\bar{V}_{n}^{\mu}-i\beta_{n}\Gamma^{\mu},\quad\hat
{A}_{n}=\bar{A}_{n}^{\mu}-\alpha_{n}u^{\mu}\label{eq:sep1}\\
\bar{V}_{n}^{\mu}  &  =\sum_{I=1}^{N_{V}}b_{n}^{I}V_{I}^{\mu},\quad\bar{A}%
_{n}^{\mu}=\sum_{I=1}^{N_{A}}a_{n}^{I}A_{I}^{\mu} \label{eq:sep2}%
\end{align}
$\allowbreak$There are $N_{V}=[(N+1)/2]$ heavy spin-1 vectors, and
$N_{A}=[N/2]$ heavy axials\footnote{$[a]\equiv$maximal integer $\leq a$.}. In
(\ref{eq:sep1}) we separated explicitly their $\Omega_{L,R}$ components
\[
\Gamma^{\mu}=\frac{1}{2i}\left(  \Omega_{R}^{\mu}+\Omega_{L}^{\mu}\right)
~,\qquad u^{\mu}=\Omega_{R}^{\mu}-\Omega_{L}^{\mu}~,
\]
which coincide with the fields already encountered in Sect. 2. The heavy mass
eigenstates transform homogeneously:%
\[
\hat{V}_{n}^{\mu}\rightarrow h\hat{V}_{n}^{\mu}h^{\dagger},\quad\hat{A}%
_{n}^{\mu}\rightarrow h\hat{A}_{n}^{\mu}h^{\dagger}%
\]
which follows from the fact that the quadratic form (\ref{eq:non-diag}) has a
zero eigenvector $\hat{V}_{0}^{\mu}=\sum_{I}\Omega_{I}^{\mu}$, and from
orthogonality of eigenvectors.

The complete Lagrangian $\mathcal{L}_{\chi}+\mathcal{L}_{\mathrm{gauge}}$ in
the form of (\ref{eq:Lgaugef}), (\ref{eq:sep}) describes in general terms an
($N+2)$-site model with arbitrary number of links, including non-sequential
ones. A particularly simple example of this general setup is the 3-site model
of Ref.~\cite{Matsuzaki:2006wn}. Here there is only one heavy vector, $\bar
{V}_{1}=V_{1}$ ($N_{A}=0$, $N_{V}=1$), with $\beta_{V}=1$ and $v_{V}=v$, and
the only free parameter of the model is $M_{V}=\sqrt{2}g_{1}v$.

It is crucial that the gauge model predicts the appearance of the commutator
term $[A_{I}^{\mu},A_{I}^{\nu}]$ in the pure gauge Lagrangian
(\ref{eq:Lgaugef}). The coefficient of this term is not determined by the
symmetry (\ref{eq:VAtrans}) and is usually set to zero in phenomenological
Lagrangians based on the massive Yang-Mills formulation
\cite{Ecker:1989yg,Bando}.

\subsection{Low-energy limit and comparison with the composite model}

To analyze the structure of the theory at low energies in the $SU(2)_{L,R}$
sector, we can express the full Lagrangian in terms of the heavy fields
$\hat{V}_{n},\hat{A}_{n}.$ The fields $V_{I}^{\mu},A_{I}^{\mu}$ have
expression of the form:%
\begin{equation}
V_{I}^{\mu}=\Gamma^{\mu}+b_{I}^{n}\hat{V}_{n},\quad A_{I}^{\mu}=\varepsilon
_{I}u^{\mu}+a_{I}^{n}\hat{A}_{n},\quad\varepsilon_{I}=a_{I}^{n}\alpha_{n},
\label{eq:subst}%
\end{equation}
which are found by inverting (\ref{eq:sep2}). The field $\Gamma^{\mu}$ enters
with unit coefficient to ensure the correct transformation properties. This
leads to:%
\begin{equation}
\mathcal{L}_{\mathrm{gauge}}=-\sum_{I\neq L,R}\frac{1}{4g_{I}^{2}}\left[
\langle\left(  \frac{1}{2}f_{+}^{\mu\nu}+\frac{i}{4}[u^{\mu},u^{\nu
}]-i\epsilon_{I}^{2}[u^{\mu},u^{\nu}]\right)  ^{2}\rangle+\langle\left(
\epsilon_{I}f_{-}^{\mu\nu}\right)  ^{2}\rangle\right]  ~+\Delta\mathcal{L}%
(\hat{V}_{n}^{\mu},\hat{A}_{n}^{\mu},u)~, \label{eq:Lshift}%
\end{equation}
where $f_{\pm}^{\mu\nu}=u\hat{W}^{\mu\nu}u^{\dagger}\pm u^{\dagger}\hat
{B}^{\mu\nu}u$.

Integrating out the heavy $\hat{V}_{n}^{\mu}$ and $\hat{A}_{n}^{\mu}$, it
follows that the interaction terms in $\Delta\mathcal{L}(\hat{V}_{n}^{\mu
},\hat{A}_{n}^{\mu},u)$ contribute to Green's functions with external
$SU(2)_{L,R}$ fields only at $\mathcal{O}(p^{6})$ (see
Ref.~\cite{Ecker:1989yg}). The only $\mathcal{O}(p^{4})$ contributions to
light-field amplitudes are the contact terms in (\ref{eq:Lshift}). In
particular, those contributing to bilinear and trilinear couplings have the
following general form
\begin{equation}
\Delta\mathcal{L}_{\mathrm{gauge}}^{(4)}=-\beta\left[  \langle(f_{+}^{\mu\nu
})^{2}\rangle+i\langle f_{+}^{\mu\nu}[u^{\mu},u^{\nu}]\rangle\right]
-\alpha\left[  \langle(f_{-}^{\mu\nu})^{2}\rangle-i\langle f_{+}^{\mu\nu
}[u^{\mu},u^{\nu}]\rangle\right]  ~,
\end{equation}
with only two (positive) free parameters: $\beta$, arising by the vector-meson
sector, and $\alpha$ arising by the axial-vector sector. For any choice of
these two couplings $\Delta\mathcal{L}_{\mathrm{gauge}}^{(4)}$ gives a
vanishing on-shell $\pi-a_{\mu}-v_{\mu}$ coupling, which contradicts the QCD
phenomenology of $\pi\rightarrow l\nu\gamma$ decays (see App.~\ref{sect:QCD}%
)\footnote{~The fact that we cannot describe the axial vectors of QCD as
massive gauge bosons has been noted long ago in Ref.~\cite{Bando}. The
so-called \emph{hidden gauge} Lagrangian Ref.~\cite{Ecker:1989yg,Bando}, which
provides a successful QCD phenomenology, is not a pure gauge Lagrangian and
differ from (\ref{eq:Lgaugef}) by the addition of trilinear couplings not
fixed by the (hidden) gauge symmetry. This problem has been recently
rediscovered in the context of 5-dimensional models: our results confirm the
problems of such models in describing the axial pion form factor~\cite{Hirn1}%
.
}.

For comparison, the $\mathcal{O}(p^{4})$ biliner and trilinear couplings of
light fields obtained integrating out the heavy vectors in the composite
models are
\begin{equation}
\Delta\mathcal{L}^{(4)}_{\mathrm{composite}} = - \frac{F_{V}^{2}}{8 M^{2}_{V}}
\langle( f_{+}^{\mu\nu})^{2} \rangle- i \frac{F_{V} G_{V}}{4 M^{2}_{V}}
\langle f_{+}^{\mu\nu} [u^{\mu},u^{\nu}] \rangle- \frac{F_{A}^{2}}{8 M^{2}%
_{A}} \langle( f_{-}^{\mu\nu})^{2}\rangle~.
\end{equation}
Comparing this result with $\Delta\mathcal{L}^{(4)}_{\mathrm{gauge}}$, we find
that the vector-meson sectors of the two models have the same structure for
$F_{V}=2G_{V}$. In the specific case of the 3-site model, requiring the same
overall normalization of the vector terms we also find the relation $F_{V} =
v$.

Contrary to the vector sector, the $\mathcal{O}(p^{4})$ structures in the
axial-vector sectors are always different in the gauge and composite
scenarios: the two models coincide only in the (trivial) case $F_{A}=\alpha=0$.

\subsection{One-loop contributions to $T$}

Without performing an explicit calculation, we can generalize the 3-site
result in (\ref{eq:Tlogs}) and show that also in the general $N$-site case
there are no large contributions to $T$ beside the infrared term.

The calculation proceeds as in Section~4. Since the effective couplings of
$B_{\mu}$ to pions and to the heavy gauge fields are only in $\mathcal{L}%
_{\chi}$, it is more convenient to avoid the field redefinitions which move
$B_{\mu}$ into $\mathcal{L}_{\mathrm{gauge}}$. The only unavoidable
redefinition is the shift $A_{I}^{\mu}\rightarrow A_{I}^{\mu}+(2\varepsilon
_{I}/v)\partial_{\mu}\pi$, necessary to obtain a canonical kinetic term for
the pions. This shift produces an effective trilinear coupling $\sim\bar
{A}_{n}\bar{V}_{m}\pi$ from the gauge Lagrangian in (\ref{eq:Lgaugef}).

At the one loop level we have at most logarithmic UV divergences. This can be
easily understood by analogy with the SM in the $m_{H}\rightarrow\infty$
limit. We are interested in the parametric dependence on the masses and
coupling constants of the coefficient of the log and of finite terms, in order
to isolate possible enhancement factors. As far as the diagrams in
Fig.~\ref{fig:T1loop} are concerned, the parametric dependence is always of
the type
\begin{equation}
\Delta\hat{T}\sim\frac{(g^{\prime})^{2}}{16\pi^{2}}\sum g_{Vn}^{2}\left(
\frac{\beta_{n}v_{Vn}^{2}}{v}\right)  ^{2}\times F_{\mathrm{loop}}\sim
\frac{(g^{\prime})^{2}}{16\pi^{2}}\sum\beta_{n}^{2}\frac{v_{Vn}^{2}}{v^{2}%
}\sim\frac{(g^{\prime})^{2}}{16\pi^{2}}~,\qquad F_{\mathrm{loop}}\sim\frac
{1}{M_{Vn}^{2}}~, \label{eq:Tgauge}%
\end{equation}
and similarly for the axial vectors. Here $M_{Vn}\sim g_{Vn}v_{Vn}$ and
$g_{Vn}$ are the effective coupling of the heavy fields obtained from
$\mathcal{L}_{\mathrm{gauge}}$. We took into account that there cannot be any
enhancement due to $\beta_{n}^{2}$ or $v_{Vn}^{2}/v^{2}$ factors: being the
matrix elements of an orthogonal rotation matrix the coefficients $|\beta
_{n}|$ are bounded by unity$;$ also, by the perturbative diagonalization of
the mass matrix one gets $\beta_{n}\sim v/v_{Vn}$ in the limit $v_{Vn}\gg v$.

The case of the diagrams in Fig.~\ref{fig:Tcutoff}, generated by the $\bar
{A}_{n}\bar{V}_{m}\pi$ trilinear coupling, is slightly more complicated.
However, after identifying the parametric dependence of the trilinear terms in
(\ref{eq:Lgaugef}) from the effective coupling of the heavy fields, one finds
also in this case $\Delta\hat{T}\sim(g^{\prime})^{2}/(16\pi^{2})$.

This general conclusion can also be reached by noting that, since in the gauge
model $M_{Vn}\sim g_{Vn}v_{Vn}$ , we cannot enhance $\Delta\hat{T}$ by a ratio
$M_{Vn}/M_{Vm}$ (as in the composite case): this would correspond to a
singularity in the limit $g_{Vn}\rightarrow\infty$ or $g_{Vm}\rightarrow0$,
while the one-loop amplitude must remain regular in this limit.

\begin{figure}[t]
\begin{center}
\includegraphics[width=18cm]{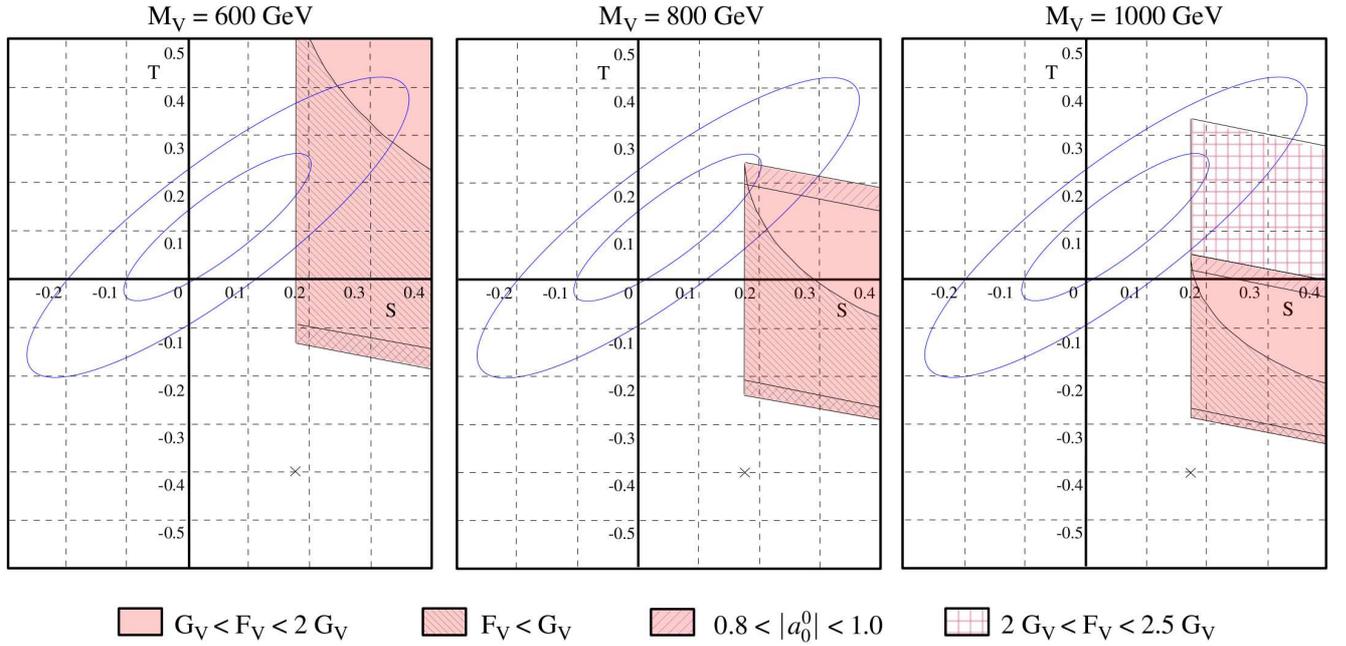}
\end{center}
\caption{Areas of $S$ and $T$ covered in the composite model for different
ranges of the free parameters.  
The two ellipses denote the experimentally
allowed region at 68\% and 95\% CL. The small cross at negative $T$ values is
the $F_{V}=F_{A}=G_{V}=0$ point (infrared logs only). }%
\label{fig:STall}%
\end{figure}

\section{Unitarity and EWPT constraints put together}
\label{sect:constr}

We can discuss now if the sole exchange of the spin-1 fields can account for
the EWPT and, in the positive case, how their properties are constrained. From
Eq.~(\ref{eq:Tlogs}) and the discussion in Sect.~\ref{sect:gauge}, we
conclude, as anticipated, that the EWPT cannot be satisfied in the gauge model
without extra, not calculable, ingredients. We thus restrict our attention to
the composite model of Sect.~\ref{sect:compo}. For our purposes the
significant constraints come solely from $S$ and $T$. The constraints from
LEP2 on $W$, $Y$ and on the cubic gauge-boson vertexes are easily satisfied.
The key equations are therefore (\ref{Tlambda2}) for $T$ and
\begin{equation}
\Delta\hat{S}=\frac{g^{2}}{96\pi^{2}}\ln\left(  \frac{\Lambda}{m_{W}}\right)
+g^{2}\left(  \frac{F_{V}^{2}}{4M_{V}^{2}}-\frac{F_{A}^{2}}{4M_{A}^{2}%
}\right)  +\ldots~, \label{eq:deltaS}%
\end{equation}
for $S$. We assume that the dots in both of these equations give negligible
contributions for $\Lambda\approx3$~TeV.\footnote{Residual finite terms are
fixed by matching $\Delta\hat{S}$ and $\Delta\hat{T}$ at $\Lambda=1$~TeV and
$F_{A}=F_{G}=G_{V}=0$, to the SM values of $\hat{S}$ and $\hat{T}$ computed
for $m_{H}=1$~TeV with the central value of $m_{top}=171.4$ GeV \cite{EWWG}.}
We also assume that below $\Lambda$ there are at most one vector and one axial
spin-1 particles. The relevant parameters are therefore $M_{V},G_{V}$ and two
dimensionless ratios $F_{V}/M_{V},F_{A}/M_{A}$.

The expectation for $S$ and $T$ in the composite model is represented in
Fig.~\ref{fig:STall} for three different values of $M_{V}$. The other
parameters are constrained by requiring that: i) the tree-level contribution
to $S$ (the second term in (\ref{eq:deltaS})) is positive; ii) $0<F_{V}%
/G_{V}<2$ or $2.5$; iii) for every $M_{V}$ the unitarity constraint on $G_{V}%
$, $|a_{0}^{0}|<1$, is satisfied. As manifest in the Figure, the saturation of
the unitarity bound is not critical. The boundary of the small $F_{V}$ region,
$F_{V}/G_{V}<1$, is also shown.

The cumulative constraints from unitarity and the EWPT are illustrated in
Fig.~\ref{fig:GVMV}, where we consider in particular two special cases,
physically distinct by the spectrum of the spin-1 particles:

\begin{enumerate}
\item[ I.] a single vector spin-1 field below $\Lambda$, denoted as
$0<F_{V}\ll2G_{V},~F_{A}=0$;

\item[II.] one gauge-like vector and one axial state below $\Lambda$, denoted
as $F_{V}=2G_{V},~F_{A}\neq0$.
\end{enumerate}

In Case I we therefore imagine that it is the quadratically divergent
\textit{vector} contribution that provides the positive $\Delta T$ necessary
to cancel the negative IR term, while the tree level contribution to $S$,
saturated by the same vector resonance, is within experimental bounds. This
may happen if $F_{V}\ll2G_{V}$. In Case II, on the other hand, the crucial
role is played by the axial spin-1, which provides the positive one-loop
$\Delta T$ and partially cancels the tree level $\Delta S$ of the vector
(which by itself would be too large). We also consider the more general case,
when both heavy vectors contribute to both $S$ and $T$:

\begin{enumerate}
\item[ III.] with one vector and one axial state below $\Lambda$, without
special constraints on $F_{V}/G_{V}$ except for the upper bound $F_{V}%
<2.5~G_{V}$, and with $F_{A}\neq0$.
\end{enumerate}

\section{Phenomenology}

\label{sect:pheno}

From a phenomenological point of view the main consequence of the previous
Section, as manifest from Fig.~\ref{fig:GVMV}, is that, unlike the pure unitarity
constraint, the EWPT require a relatively light spin-1 vector. In turn,
together with unitarity, this bounds $G_{V}$ between about 120 and 180~GeV.
Less is known about the axial spin-1, since it does not influence the $\pi\pi
$, or $W_{L}W_{L}$, scattering amplitude and it enters the EWPT only through
the combination $F_{A}/M_{A}$.

In the allowed region of the $(M_{V}, G_{V})$ plane the cross section for
$\sigma(pp \to V \to W Z + 2 \mathrm{jets})$ at the LHC via vector boson
fusion, taking into account the leptonic branching ratios of $W$ and $Z$, and
applying the cuts suggested in Ref.~\cite{Belyaev} on the jet variables, is
between 0.3 and 3 fb. The coupling of the heavy vector to the fermions
proceeds only through its mixing with the $W$ and the $Z$ and is therefore
highly suppressed. As a consequence the production by quark-antiquark
annihilation is irrelevant, as is probably irrelevant also the search of the
neutral vector decaying into a pair of charged leptons.

The relevant widths for $g^{\prime}=0$ and up to corrections of order
$m_{W}^{2}/M_{V}^{2}$ are
\begin{equation}
\Gamma(V^{+}\rightarrow W_{L}^{+}Z_{L})\approx\Gamma(V^{0}\rightarrow
W_{L}^{+}W_{L}^{-})\approx\frac{G_{V}^{2}M_{V}^{3}}{48\pi v^{4}}%
\approx 9~\mathrm{GeV}\left(  \frac{G_{V}}{150~\mathrm{GeV}}\right)
^{2}\left(  \frac{M_{V}}{600~\mathrm{GeV}}\right)  ^{3}~,
\end{equation}
and
\begin{equation}
\Gamma(V^{0}\rightarrow\ell^{+}\ell^{-})=\frac{g^{4}F_{V}^{2}}{384\pi M_{V}}~,
\end{equation}
so that
\begin{equation}
\frac{\Gamma(V^{0}\rightarrow\ell^{+}\ell^{-})}{\Gamma(V^{0}\rightarrow
W_{L}^{+}W_{L}^{-})}\approx\frac{g^{4}}{8}\frac{F_{V}^{2}}{G_{V}^{2}}%
\frac{v^{4}}{M_{V}^{4}}\approx 6\cdot10^{-4}\frac{F_{V}^{2}}{G_{V}^{2}}\left(
\frac{600~\mathrm{GeV}}{M_{V}}\right)  ^{4}.
\end{equation}

Notice the rather narrow total width of $V$, compared to the width of the SM
Higgs boson of the same mass. This fact can be partially explained by the fact
that: i) \textit{three} vectors rather than \textit{one} scalar boson unitarize
the $W_{L}W_{L}$ scattering, so that the unitarization is achieved for
relatively lower values of the relevant cubic coupling;
ii) the vectors decay in to $W_{L}W_{L}$ pairs in  $P$ wave
(the unitarization of $S$-wave  $W_{L}W_{L}$ scattering occurs via $t$-channel
exchange) contrary to $S$-wave decay of the Higgs boson.

The $V^{+}$ production at LHC is supposedly easier to search for, compared to
the $V^{0}$, because of the relatively easier reconstruction of the $W^{+}Z$
versus the $W^{+}W^{-}$ final state. Generalizing the results of
Ref.~\cite{Belyaev} to our framework, with 100 fb$^{-1}$ one should expect a
clear signal (more than 5$\sigma$) for the $V^{+}$ in the whole allowed region
in Fig.~\ref{fig:GVMV} (for related phenomenological studies see
also~\cite{others}). As already mentioned, the searches at LEP2 both in the
$l^{+}l^{-}$ and in the $W^{+}W^{-}$ channels do not add any significant
constraint on Fig.~\ref{fig:GVMV} even in the low mass region.

\section{Summary and conclusions}

As said at the beginning, the \textit{double} role played by a relatively
light Higgs boson in allowing a consistent description of the EWPT in the SM
and in unitarizing the $W_{L}W_{L}$ scattering speaks strongly in favour of
its discovery at the LHC. Any alternative point of view has a hard time in
replacing this role in a convincingly calculable (rather than \textit{ad hoc})
way. In this work we have analyzed up to which point one or more spin-1
particles can be such an alternative.

While the presence of and the motivation for spin-1 particles emerges in
different contexts, we have focussed on their phenomenological description. To
this end, building on the much suggestive $SU(2)_{L}\times SU(2)_{R}%
/SU(2)_{L+R}$ symmetry, we have considered two different models:

\begin{itemize}
\item A \textit{composite} model, inspired by (but also departing from) QCD,
where the heavy bosons are triplet representations of $SU(2)_{L}\times
SU(2)_{R}/SU(2)_{L+R}$;

\item A \textit{gauge} model where $SU(2)_{L}\times SU(2)_{R}/SU(2)_{L+R}$ is
extended to $SU(2)^{N}/SU(2)_{diag}$ and the heavy vectors emerge from the
breaking of the extended gauge group.
\end{itemize}

Both model are described by effective Lagrangians with a cutoff $\Lambda
\approx4\pi v\approx3~$TeV. Each Lagrangian would support counterterms
associated with the EWPT parameters $S$ and $T$. To defend the calculability
of the EWPT we have to take them negligible.

We have argued that the gauge model is a special case of the composite model,
while the composite model cannot be reduced to the gauge model except for
special values of the parameters. As known for example from work in QCD, only
the composite model can account for some phenomenological properties of pion
dynamics. This difference proves in fact essential to offer a potentially
consistent description of the EWPT in terms of the sole exchange of the heavy
vectors in the composite model.

The problem of the $S$-parameter in Higgs-less models, positive and
potentially too large, is well known. Its compatibility with data depends
crucially, however, on its correlation with $T$, which receives a large
negative \textit{infrared} effect. $S$ and $T$ must therefore be discussed
together, which requires the inclusion of the full $T$ at one loop. In the
composite model, the crucial effect in the $T$-parameter comes from the loop
diagram in Fig.~\ref{fig:T1loop}b with derivative couplings to the Goldstone
field of the heavy spin-1 particles. This is the only quadratically divergent
diagram. As such, for light enough spin-1 fields, it can compensate the
logarithmically divergent \textit{infrared} effect which does not involve the
exchange of any heavy particle, as illustrated in Fig.~\ref{fig:STall}. If we
impose that suitable correlation functions among composite states have good
ultraviolet behavior, the cutoff $\Lambda$ in the quadratically divergent
diagram gets replaced by a heavy vector mass. We assume this mass to be close
to the cutoff, so that the naive estimate of $T$ with the cutoff itself
continues to be reasonable.

On this basis, the EWPT constraints can be crossed with the ones coming from
unitarity in longitudinal gauge-boson scattering at high energy. The
corresponding significant restriction in the parameter space of the lightest
vector spin-1 state is summarized in Fig.~\ref{fig:GVMV}. Note how the
unitarity constraint by itself would not give a strong upper bound on $M_{V}$.
This is a crucial feature in connection with the searches at the LHC, as we
comment in Sect. \ref{sect:pheno}.

It goes without saying that we do not know if a phenomenological model like
this can emerge from any fundamental theory of electroweak symmetry breaking.
In particular we are sharply aware of the fact that this has left us with many
free parameters to play with. As repeatedly said, it is hard to compete with
the Higgs boson picture of electroweak symmetry breaking. Yet we find it
interesting that there be a plausible mechanism that can account both for the
EWPT and the unitarity in terms of heavy spin-1 exchanges and that, under
suitable hypotheses, this can constrain in a significant way the properties of
the lightest of them, most likely accessible to LHC searches.

\section*{Acknowledgements}

V.S.R. would like to thank Roberto Contino for useful discussions. R.B. and
V.S.R. were partially supported by the EU under RTN contract
MRTN-CT-2004-503369. R.B. was also supported in part by MIUR 
under the contract PRIN-2006022501.
V.S.R. was also partially supported by the EU under ToK
contract MTKD-CT-2005-029466; he would like to acknowledge the hospitality of
the Institute of Theoretical Physics of Warsaw University, where a part of
this work was carried out. G.I.~was partially supported by the EU
under RTN contract MTRN-CT-2006-035482.

\appendix

\section{High momentum behavior of two and three point functions}

\label{sect:sum}

The consistency of the theory requires a smoothing in the ultraviolet of the
correlation functions involving external SM gauge fields or, more generally,
external vector or axial sources of $SU(2)_{L} \times SU(2)_{R}$, with respect
to the potential bad behaviour inferred from the lowest-order chiral
Lagrangians. This implies a series of sum rules for the free parameters of the model.

Starting from the two-point functions,
\begin{equation}
i\int d^{4}xe^{iqx}[\langle0|v_{i}^{\mu}(x)v_{j}^{\nu}(0)-a_{i}^{\mu}%
(x)a_{j}^{\nu}(0)|0\rangle]=\delta_{ij}g^{\mu\nu}q^{2}\Pi_{LR}(q^{2}%
)+\mathcal{O}(q^{\mu}q^{\nu})~,
\end{equation}
the cancellation of tree-level contributions to $q^{4}\Pi_{LR}(q^{2})$ at high
$q^{2}$ leads to the well-known Weinberg sum rules,
\[
\sum_{i=1}^{N_{V}}F_{V_{i}}^{2}-\sum_{i=1}^{N_{A}}F_{A_{i}}^{2}=v^{2}%
~,\qquad\quad\sum_{i=1}^{N_{V}}F_{V_{i}}^{2}M_{V_{i}}^{2}-\sum_{i=1}^{N_{A}%
}F_{A_{i}}^{2}M_{A_{i}}^{2}=0~,
\]
for a generic model with $N_{V}$ vector and $N_{A}$ axial fields (with $F_{V}$
and $G_{V}$ defined as in Sect.~\ref{sect:compo}).

In the case of three point functions, from the pion vector form factor,
\begin{equation}
\langle\pi(p^{\prime})|v^{\mu}(q)|\pi(p)\rangle=F_{\pi}^{v}(q^{2}%
)(p+p^{\prime})^{\mu}~,
\end{equation}
the condition $F_{\pi}^{v}(q^{2})\rightarrow0$ implies
\begin{equation}
\sum_{i}F_{V_{i}}G_{V_{i}}=v^{2}~. \label{VMD}%
\end{equation}
Similarly, defining
\begin{align}
\langle A(p^{\prime},\epsilon)|v^{\mu}(q)|\pi(p)\rangle &  =F_{A\pi}^{v}%
(q^{2})[(pq)\epsilon_{\mu}-q^{\mu}(p\epsilon)]+\mathcal{O}[(p^{\prime}%
)^{2}]~,\\
\langle V(p^{\prime},\epsilon)|a^{\mu}(q)|\pi(p)\rangle &  =F_{V\pi}^{a}%
(q^{2})[(pq)\epsilon_{\mu}-q^{\mu}(p\epsilon)]+\mathcal{O}[(p^{\prime})^{2}]~,
\end{align}
the conditions $F_{A\pi}^{a}(q^{2})\rightarrow0$ and $F_{V\pi}^{v}%
(q^{2})\rightarrow0$ lead to
\begin{equation}
2\sum_{j=1}^{N_{V}}g_{j}^{A_{i}}F_{V_{j}}=F_{A_{i}}~,\qquad2\sum_{j=1}^{N_{A}%
}g_{j}^{V_{i}}F_{A_{j}}=F_{V_{i}}-2G_{V_{i}}~, \label{eq:sumr3}%
\end{equation}
with $g_{j}^{A}$ and $g_{j}^{V}$ defined as in Eq.~(\ref{eq:L2V}). These last
two conditions ensure the cancellation of the quadratic divergence in $T$ of
the composite model.

In the gauge models some of these sum rules are satisfied in a trivial way.
For instance, in the 3-site model all terms in Eq.~(\ref{eq:sumr3}) are
identically zero. In the more general case of composite models we expect the
sum rules to be satisfied summing over the first few sets of resonances (not
necessarily a single set). With two, non-degenerate sets of vector and
axial-vector states, one light and the other close to the cut-off, all the sum
rules can be satisfied choosing arbitrary values for the light parameters
$M_{V}$, $F_{V}$, $G_{V}$, and $F_{A}/M_{A}$, within the ranges discussed in
Sect.~\ref{sect:constr}.

\section{Heavy spin-1 resonances in QCD}

\label{sect:QCD}

Low-energy QCD is a theory with spontaneously broken approximate
$SU(3)_{L}\times SU(3)_{R}$ global symmetry. For comparison with the EWSB we
will ignore the heavier $s$ quark and consider the $SU(2)_{L}\times SU(2)_{R}$
subgroup. In this Appendix we collect experimentally measured parameters of
spin-1 resonances in QCD. If the EWSB were described by a QCD-like theory, all
the parameters with dimension of mass would have to be scaled up by the ratio
$v/F_{\pi}$, where the pion-decay constant%
\[
F_{\pi}\simeq93.3\,\text{MeV}%
\]
is the QCD scale of $SU(2)_{L}\times SU(2)_{R}$ breaking.

The lightest vector and axial spin-1 QCD resonances are the $\rho\equiv V$ and
the $a_{1}\equiv A$:%
\begin{align*}
M_{\rho}  &  \simeq775\text{ MeV, \quad}\Gamma_{\rho}\simeq150\text{ MeV,}\\
M_{a1}  &  \simeq1230\text{ MeV,\quad}\Gamma_{a1}=250\div600\text{
MeV\thinspace.}%
\end{align*}
The $\rho$ width is totally dominated by $\rho\rightarrow\pi\pi$, which allows
the determination of $G_{V}$. From the tree-level amplitude we get
\begin{equation}
G_{V}\simeq60\text{ MeV}~(M_{\pi}=0) \qquad\mathrm{or}\qquad G_{V}%
\simeq67\text{ MeV}~(M_{\pi}=135~\text{MeV})~. \label{eq:GV-QCD}%
\end{equation}
Note that as long as we neglect light quark masses (or explicit chiral
symmetry breaking terms), we have no reasons to prefer the value of $G_{V}$
evaluated in the chiral limit ($M_{\pi}=0$), versus the one obtained with the
physical pion mass.

If rescaled to the electroweak scale, the $\rho$ meson would correspond to a
$2$ TeV spin-1 boson. The rescaled $G_{V}$ is at the lower limit of the
unitarity band in Fig.~\ref{fig:GVMV}, in agreement with the hypothesis that
the $\rho$-meson plays a key role in unitarizing $\pi\pi$ scattering in QCD.

The $F_{V}$ and $F_{A}$ are determined from $\Gamma(\rho\rightarrow e^{+}%
e^{-})$ and $\Gamma(a_{1}\rightarrow\pi\gamma)$,
respectively~\cite{Ecker:1989yg}:\footnote{Notice that \cite{Ecker:1989yg}
used a different value of $M_{a1}$.}
\begin{align}
F_{V}  &  =157\text{ MeV}~,\nonumber\\
F_{A}  &  =(120\pm25)\text{ MeV}~.\label{eq:FV-QCD}%
\end{align}
Finally, the coupling $g^{A}$ can be determined from $\Gamma(a_{1}%
\rightarrow\rho\pi)\simeq60\%\Gamma_{a1}$ \cite{PDG}. Since the gauge model
relation $F_{V}=2G_{V}$ is satisfied at a $20\%$ level, it is reasonable to
assume that $g^{V}\ll1$ (see Eq.~(\ref{eq:sumr3})), so that the first operator
in (\ref{eq:L2V}) dominates this width. Under this assumption, we extract:
\[
g^{A} = 0.50\pm0.15~\text{.}%
\]

It is now interesting to see if the above resonance parameters, extracted from
experiment, satisfy the remaining sum rules of Appendix~\ref{sect:sum}. Both
Weinberg sum rules are satisfied within the experimental errors, although the
favored values of $F_{V}$ are different, which indicates potential importance
of higher-mass resonances. The Vector Meson Dominance relation for the pion
form factor (\ref{VMD}) is well satisfied by the sole $\rho$ contribution if
we choose $G_{V}$ in the lower range of (\ref{eq:GV-QCD}), indicating a small
but non negligible breaking of the $F_{V}=2G_{V}$ relation. The first of the
two sum rules (\ref{eq:sumr3}) is also in good agreement for the above $g^{A}%
$. This means that in QCD the quadratic divergence in the charged-neutral pion
wavefunction renormalization difference, present due to $F_{A}\neq0$ (see
Eq.~(\ref{Tlambda2})), is likely cut off at a scale of $M_{\rho}$ ($\sim2$ TeV
in electroweak units). Similarly, we can expect the contribution of the
quadratic divergence proportional to $F_{V}-2G_{V}$ to be cut off by the
$M_{a_{1}}$ mass ($\sim3$ TeV in electroweak units).

In principle, another interesting source of information is provided by the
axial pion form factor, or the $\pi-v_{\mu}(q)-a_{\nu}(k)$ coupling at
$q^{2}= k^{2}=0$, measured with high precision in $\pi^{+}\rightarrow e^{+}%
\nu\gamma$~\cite{PIBETA}:
\bea 
\langle \gamma(\epsilon,q)| \bar u \gamma_\mu\gamma_5 d | \pi(p) \rangle 
&=& \frac{i e}{M_\pi} \mathcal{F}_{A} \left[ (p q) \epsilon_\mu - (p\epsilon) q_\mu \right] + \text{pion pole term} \\
\mathcal{F}_{A}^{\mathrm{exp}}&=&0.0115\pm0.0004 ~.\label{eq:FAexp}%
\eea
As stressed in Sect.~\ref{sect:T} and~\ref{sect:gauge}, the non-vanishing of
$\mathcal{F}_{A}$ is a clear evidence for $F_{A}\neq0$ and/or $F_{V}\neq
2G_{V}$. Indeed the $\mathcal{O}(p^{4})$ value of $\mathcal{F}_{A}$ obtained
by spin-1 resonance exchange is \cite{Ecker:1989yg}%
\begin{equation}
\mathcal{F}_{A}^{\mathrm{vectors}} = \frac{\sqrt{2}M_{\pi}}{F_{\pi}}\left(
\frac{F_{A}^{2}}{M_{A}^{2}}-\frac{F_{V}(F_{V}-2G_{V})}{M_{V}^{2}}\right)
~.\label{eq:FApred}%
\end{equation}
Since axial and vector contributions in (\ref{eq:FApred}) turn out to have
opposite sign, $\mathcal{F}_{A}^{\mathrm{exp}}$ does not provide, by itself, a
clear independent determination of $F_{A}$ and $F_{V}-2G_{V}$. However, it
provides a useful cross-check of the whole picture: using $F_{V}$ and $F_{A}$
from (\ref{eq:FV-QCD}) as input values, one extracts $G_{V} = (70 \pm10)$~MeV,
in good agreement with (\ref{eq:GV-QCD}).


\begin{thebibliography}{99}                                                                                               %
\bibitem {Casalbuoni:1985kq}R.~Casalbuoni, S.~De Curtis, D.~Dominici and
R.~Gatto,
Phys.\ Lett.\ B \textbf{155} (1985) 95; Nucl.\ Phys.\ B \textbf{282} (1987) 235.

\bibitem {Csaki:2003dt}C.~Csaki, C.~Grojean, H.~Murayama, L.~Pilo and
J.~Terning,
Phys.\ Rev.\ D \textbf{69}, 055006 (2004) [arXiv:hep-ph/0305237].

\bibitem {Foadi:2003xa}R.~Foadi, S.~Gopalakrishna and C.~Schmidt,
JHEP \textbf{0403} (2004) 042 [arXiv:hep-ph/0312324].


\bibitem {Georgi:2004iy}H.~Georgi,
Phys.\ Rev.\ D \textbf{71} (2005) 015016 [arXiv:hep-ph/0408067].

\bibitem {Ecker:1989yg}G.~Ecker \emph{et al.},
Phys.\ Lett.\ B \textbf{223} (1989) 425;
Nucl.\ Phys.\ B \textbf{321} (1989) 311.




\bibitem {coleman}S.~R.~Coleman \textit{et. al.} Phys.\ Rev.\ \textbf{177},
2239, 2247 (1969);
C.G.~Callan, \textit{et. al.}
Phys.\ Rev.\ \textbf{177} (1969) 2247.
For recent reviews, see e.g.:~G.~Ecker,
Prog.\ Part.\ Nucl.\ Phys.\ \textbf{35}, 1 (1995) [arXiv:hep-ph/9501357];
G.~Colangelo and G.~Isidori,
arXiv:hep-ph/0101264.


\bibitem {Bagger}J.~Bagger~\emph{et al.,} Phys.\ Rev.\ D \textbf{49} (1994) 1246.

\bibitem {Papucci}M.~Papucci,
arXiv:hep-ph/0408058.


\bibitem {Matsuzaki:2006wn}S.~Matsuzaki, R.~S.~Chivukula, E.~H.~Simmons and
M.~Tanabashi,
Phys.\ Rev.\ D \textbf{75} (2007) 073002 [arXiv:hep-ph/0607191].

\bibitem {Lee}B.~W.~Lee, C.~Quigg and H.~B.~Thacker,
Phys.\ Rev.\ D \textbf{16}, 1519 (1977).


\bibitem {Barbieri:2004qk}R.~Barbieri, A.~Pomarol, R.~Rattazzi and
A.~Strumia,
Nucl.\ Phys.\ B \textbf{703} (2004) 127 [arXiv:hep-ph/0405040].


\bibitem {Barbieri:1992dq}R.~Barbieri, M.~Beccaria, P.~Ciafaloni, G.~Curci and
A.~Vicere,
Nucl.\ Phys.\ B \textbf{409} (1993) 105.


\bibitem {PIBETA}E.~Frlez \textit{et al.},
Phys.\ Rev.\ Lett.\ \textbf{93} (2004) 181804 [arXiv:hep-ex/0312029].


\bibitem {Barbieri:2003pr}R.~Barbieri, A.~Pomarol and R.~Rattazzi,
Phys.\ Lett.\ B \textbf{591} (2004) 141 [arXiv:hep-ph/0310285].


\bibitem {Bando}M.~Bando, T.~Fujiwara and K.~Yamawaki,
Prog.\ Theor.\ Phys.\ \textbf{79} (1988) 1140.


\bibitem {Hirn1}J.~Hirn and V.~Sanz,
JHEP \textbf{0512} (2005) 030 [arXiv:hep-ph/0507049].


\bibitem {EWWG} www.cern.ch/LEPEWWG

\bibitem {Belyaev}H.~J.~He \textit{et al.},
arXiv:0708.2588 [hep-ph];
A.~Belyaev,
arXiv:0711.1919 [hep-ph].


\bibitem {others}A.~Birkedal, K.~Matchev and M.~Perelstein,
Phys.\ Rev.\ Lett.\ \textbf{94}, 191803 (2005) [arXiv:hep-ph/0412278].
G.Azuelos, P-A. Delsart, and J. Id\'{a}rraga, in
arXiv:hep-ph/0602198, p.235.
M.~Fabbrichesi and L.~Vecchi,
Phys.\ Rev.\ D \textbf{76}, 056002 (2007) [arXiv:hep-ph/0703236].


\bibitem {PDG}W.-M.Yao et al. (Particle Data Group), J. Phys. G 33, 1 (2006)
and 2007 partial update for the 2008 edition



\end{thebibliography}
\end{document}